# Importance of Paramagnetic Background Subtraction for Determining the Magnetic Moment in Epitaxially Grown Monolayer and Few-Layer van der Waals Magnets

Dante J. O'Hara[1*], Tiancong Zhu[2*], and Roland K. Kawakami[1,2]

[1] Materials Science and Engineering, University of California, Riverside, CA 92521
[2] Department of Physics, The Ohio State University, Columbus, OH 43210
* Authors contributed equally

*Abstract*—Due to the atomically thin nature of monolayer and few-layer van der Waals magnets, the undesired background signal from the substrate can have significant contribution when characterizing their magnetic properties. This brings challenges in accurately determining the magnitude of the magnetic moment of the epitaxially grown van der Waals magnets on bulk substrates. In this paper, we discuss the impact of the background subtraction method for accurately determining the magnetic moments in such systems. Using the recently reported intrinsic two-dimensional (2D) van der Waals ferromagnet $MnSe_2$ as an example, we show that a normal diamagnetic background subtraction method in analyzing the bulk magnetometry measurement will result in an unexpectedly large magnetic moment (greater than ~10 $\mu_B$ per formula unit). Through our systematic growth study, we identify an additional paramagnetic signal due to unintentional Mn doping of the substrate. To extract the correct magnetic moment, a paramagnetic background should also be considered. This yields a total magnetic moment of ~4 $\mu_B$ per formula unit in monolayer $MnSe_2$, which is in close agreement to the theoretically predicted value.

*Index Terms*—Nanomagnetics, 2D van der Waals magnet, ferromagnetism, molecular beam epitaxy

## I. INTRODUCTION

Realizing and understanding magnetism in two-dimensional (2D) materials has been a fascinating topic for physicists for many decades. Although extensively studied theoretically [Onsager 1944, Mermin 1966, Sivadas 2015], the experimental demonstration of ferromagnetic ordering in 2D materials was not achieved until 2017 [Gong 2017, Huang 2017]. Using the mechanical exfoliation method on van der Waals $CrI_3$ and $Cr_2Ge_2Te_6$, researchers thinned the crystals down to monolayers (bilayers for $Cr_2Ge_2Te_6$) and obtained ferromagnetic signals from µm-sized flakes at cryogenic temperatures. Since these discoveries, extensive studies have been performed on both materials, and exciting properties, such as gate tunable magnetism [Wenyu 2017, Huang 2018, Jiang 2018a, Jiang 2018b, Wang 2018c], strong magnetic proximity when coupled to a non-magnetic material [Zhong 2017], giant tunneling magnetoresistance and spin-filtering effects [Kim 2018, Klein 2018, Song 2018, Wang 2018a] have been reported. Subsequently, other 2D van der Waals magnets, such as $Fe_3GeTe_2$, have also been discovered and studied extensively via mechanical exfoliation from a bulk single crystal [Deng 2018, Fei 2018, Wang 2018b]. Another breakthrough in the development of 2D magnets is the epitaxial growth of monolayer van der Waals magnets in the transition metal dichalcogenide (TMDC) family. By using molecular beam epitaxy (MBE), Bonilla, *et al.* and O'Hara, *et al.* have separately reported ferromagnetism in large area van der Waals materials, $VSe_2$ [Bonilla 2018] and $MnSe_2$ [O'Hara 2018] down to the monolayer limit. Remarkably, the ferromagnetism in both $VSe_2$ and $MnSe_2$ persists up to and above room temperature, which is crucial for industrial applications in magnetic memory. These advances in 2D van der Waals magnets show great promise for future information storage and non-volatile logic technologies.

The ability to synthesize large area van der Waals magnets also opens the possibility to measure their magnetic properties via bulk magnetometry methods, such as superconducting quantum interference device (SQUID) and/or vibrating sample magnetometry (VSM). However, challenges have also arisen when characterizing the 2D magnets with the above techniques. Due to the atomically thin nature of the monolayer and few-layer van der Waals magnets, the undesired magnetic background signal from the substrate can have significant contribution to the total magnetization measurement. This brings difficulty in accurately determining the magnetic moment in these materials. For example, an extremely large magnetic moment of ~15 $\mu_B$ per formula unit was reported for a monolayer of $VSe_2$ grown on highly oriented pyrolytic graphite (HOPG) and $MoS_2$ [Bonilla 2018], which exceeds the theoretically predicted value of ~0.7 $\mu_B$ per formula unit [Ma 2012]. On the other hand, the magnetic moment of ~4 $\mu_B$ per formula unit reported for a monolayer of $MnSe_2$ grown on GaSe(0001) [O'Hara 2018] is in agreement with density functional theory (DFT) calculations [Ataca 2012, Kan 2014]. Such discrepancies in the experimentally obtained total magnetic moment can be due to different background subtraction methods when analyzing the magnetization results, where different contributions of the background signal are considered. The choice of the background subtraction method can strongly affect the total magnetic moment extracted from the experimental results.

In this Letter, we use epitaxially grown $MnSe_2$ monolayers on GaSe(0001)/GaAs(111)B substrates as an example to demonstrate how the background subtraction method can affect the value of the magnetic moment extracted from the experimental measurement. Importantly, systematic control measurements identify the presence of a paramagnetic signal that originates from unintentional Mn impurities in the substrate, most likely incorporated in the GaSe layer during synthesis. When the background subtraction only accounts for the diamagnetism from the substrate and ignores the presence of the paramagnetic signal, we obtain unexpectedly large magnetic moment values greater than ~10 $\mu_B$ per Mn. On the other hand, when the background subtraction includes both the diamagnetic and paramagnetic components, we obtain magnetic moment values of ~4 $\mu_B$ per Mn, which is consistent with DFT calculations [Ataca 2012, Kan 2014]. Analysis of the background subtraction procedure including the paramagnetic component demonstrates its improved reliability and accuracy as compared to typical procedures only considering the diamagnetic component. Moreover, our results show that a careful background subtraction is crucial for further study of monolayer and few-layer van der Waals magnets grown by MBE on bulk substrates.

## II. EXPERIMENTAL METHODS

The $MnSe_2$ samples are prepared by van der Waals epitaxy [Walsh 2017] in a Veeco GEN930 MBE chamber on GaSe(0001)/GaAs(111)B substrates following the recipe of O'Hara, *et al.* in a recent report [O'Hara 2018] with a base pressure of $2\times10^{-10}$ Torr. Epi-ready, un-doped GaAs(111)B substrates (AXT, single-side polished, 0.5 mm thick, 0° ± 0.5° offcut, $1.4\times10^8$ Ω-cm resistivity) are indium-bonded to an unpolished Si backing wafer and annealed under a Se flux (beam equivalent pressure (BEP) of ~$2\times10^{-6}$ Torr) at 600°C for 20 min under ultra-high vacuum to remove the surface oxide and terminate the surface with Se. The sample is then cooled to a substrate temperature of 400°C for the base layer growth of GaSe. The growth is monitored in real-time via reflection high-energy electron diffraction (RHEED) at an operating voltage of 15 kV. The substrate temperature is measured using a thermocouple that is attached to the continuous azimuthal rotation (CAR) manipulator substrate heater. Standard Knudsen-style effusion cells are used for the deposition of Ga (United Mineral & Chemical Corporation, 99.99999%) and Mn (Alfa Aesar, 99.98%) with typical cell temperatures of 1000°C and 800°C, respectively, while a valved cracking source is operated at 950°C (bulk zone, 290°C) for the deposition of atomic Se (United Mineral & Chemical Corporation, 99.9999%). The growth is performed under a Se overpressure with a BEP flux ratio of ~100 for Se:Ga and ~60 for Se:Mn, where the Se re-evaporates. The beam fluxes are measured using a nude ion gauge with a tungsten filament positioned for growth and the corresponding deposition rate is calibrated based on film thicknesses determined by x-ray reflectometry (Bruker, D8 Discover). The typical deposition rate is 1 nm/min for GaSe and 0.1 ML/sec for $MnSe_2$. Samples are capped with amorphous selenium (a-Se) at room temperature prior to removal from the chamber to protect the surface from oxidation and degradation. The lattice structure for monolayer $MnSe_2$ and the heterostructure for the samples discussed in this study are shown in Figures 1a and 1b, respectively.

The GaSe base layer growths are followed by two different growth recipes for van der Waals $MnSe_2$. The first keeps the sample temperature at 400°C while the second lowers the sample temperature to 300°C for the deposition of $MnSe_2$. The former is employed by closing the Ga shutter (while the Se shutter remains open) and opening the Mn shutter immediately. This provides streakier RHEED patterns up to high thicknesses (> 40 nm) and transitions to antiferromagnetic α-MnSe(111) according to x-ray diffraction (XRD) measurements, shown in Figure 1c. The latter is employed by closing both the Ga and Se shutters and cooling the sample to 300°C. Once the sample temperature is stable, the Se and Mn shutters are opened to deposit $MnSe_2$ layers. This recipe leads to rougher films according the spotty RHEED patterns when the sample gets thicker (> 5 MLs) [O'Hara 2018]. XRD scans show identical peaks to the 400°C growth, shown in Figure 1c. While the results of this paper apply for both types of samples, the representative analysis is presented for samples grown at 400°C.

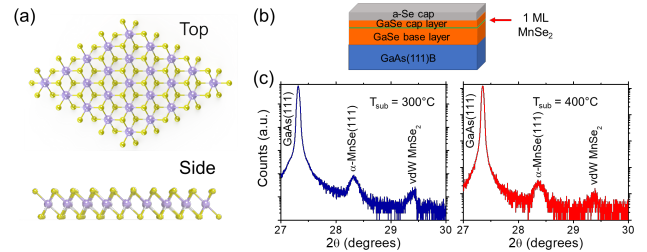

Figure 1. Lattice structure and properties of monolayer 1T-$MnSe_2$. (a) Ball-and-stick model showing top and side views of monolayer 1T-$MnSe_2$. Purple represents Mn and yellow represents Se. (b) Schematic showing the $MnSe_2$/GaSe(0001) heterostructure grown on a GaAs(111)B substrate. (c) XRD characterization of 40 nm $MnSe_x$ films grown on GaSe(0001)/GaAs(111)B substrates showing additional peaks corresponding to α-MnSe(111) and 1T-$MnSe_2$.

Superconducting quantum interference device (SQUID) magnetometry (Quantum Design, MPMS XL) measurements are used to measure the magnetic properties of the samples. Samples are mounted in the out-of-plane orientation (with the magnetic field applied perpendicular to the film surface) and are measured at room temperature.

## III. RESULTS AND DISCUSSION

Monolayer $MnSe_2$ films are grown on 55 nm GaSe(0001)/GaAs(111)B and are capped with 5 nm GaSe and amorphous Se before removing the sample from the MBE chamber (schematic shown in Figure 1b). Room temperature, out-of-plane magnetization loops show magnetic hysteresis indicating ferromagnetic ordering for 1 ML $MnSe_2$ on the GaSe base layer. Figure 2a shows the SQUID magnetization loop without any background subtracted and is normalized to the area of the sample.

The 1 ML $MnSe_2$ hysteresis loop shows an obvious linear diamagnetic background at high magnetic field, which arises from the bulk GaAs(111) substrate. The standard data analysis procedure for extracting the ferromagnetic signal is using a linear background subtraction to remove the diamagnetic signal from the substrate. Figure 2b shows the magnetization loop after applying a linear fit to the diamagnetic background in Figure 2a and subtracting out the signal. Although the hysteresis loop clearly closes at low magnetic field (~2 kOe), the magnetic moment does not saturate until higher fields (~15 kOe), indicating a possible paramagnetic component. In addition, the saturation moment at room temperature is calculated to be ~12 $\mu_B$/Mn (Figure 2b, inset) after subtracting the linear background, which is approximately three times as large than what is predicted in DFT calculations [Kan 2014]. The strong and non-saturating magnetic moment at high magnetic field draws attention for understanding the contributions to the bulk magnetization loop.

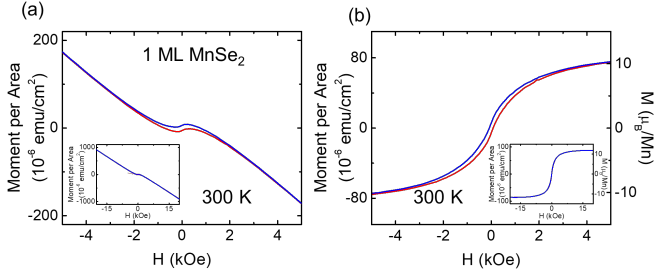

**Figure 2. Room temperature, out-of-plane magnetic hysteresis loop of 1 ML MnSe$_2$. (a)** Raw bulk magnetization loop showing diamagnetic background from substrate. Inset: Full range M(H) scan up to higher magnetic fields. **(b)** Bulk magnetization loop after linear diamagnetic background subtraction of raw data. Inset: Full range M(H) scan up to higher magnetic fields.

An interesting question is to understand what is contributing to the large background signal in the 1 ML MnSe$_2$ SQUID magnetization loop. One possibility is that the background signal comes from the GaSe(0001) base layer. It is worthwhile mentioning that during the annealing of the GaAs substrate and growth of the GaSe base layer, the Mn effusion cell is at its deposition temperature ($T_{Mn}$ = 800°C). Although the Mn effusion cell is masked by a shutter during this process, trace amounts of Mn impurities might possibly bypass the shutter and incorporate into the base layer and/or substrate due to the relatively high vapor pressure of Mn. To examine this case, we systematically study the GaSe base layer growth under identical conditions and characterize its magnetic properties.

To investigate the contribution of the GaSe base layer to the magnetic signal, we grow a 60 nm GaSe(0001)/GaAs(111)B sample under identical conditions to the 1 ML MnSe$_2$ sample with the Mn cell maintained at its deposition temperature of 800°C (we label this as: 0 ML "Mn hot"). The magnetic properties of this sample show a non-linear signal which has a large saturation magnetic moment of ~4.3×10$^{-5}$ emu/cm$^2$ with no remanent magnetization after the diamagnetic background subtraction (Figure 3b). There is no MnSe$_2$ deposited in this case and the saturation magnetic moment has similar order of magnitude to what is shown in Figure 2, so the paramagnetic-like signal should only come from the substrate. We further employ a Brillouin fitting on the extracted curve, which is shown in the bottom panel in Figure 3a. The Brillouin fitting agrees very well with the observed non-linear signal, which confirms the paramagnetic signature in this sample.

The Mn impurities that are potentially incorporating inside the Ga$_{1-x}$Mn$_x$Se matrix may contain randomly oriented magnetic moments that have no direct exchange interaction with each other [Pekarek 1998]. This can give rise to a paramagnetic contribution in the magnetization signal which can be explained by the Brillouin function. To further confirm that the paramagnetic signal is coming from Mn doping in the GaSe(0001) base layer, we grow a nominally identical 0 ML control sample but with the Mn effusion cell temperature lowered to an idle state ($T_{Mn}$ = 620°C, which we label as: 0 ML "Mn cold"). By lowering the temperature of the effusion cell, the Mn vapor pressure should decrease leading to less impurities in the base layer and/or substrate. Indeed, after measuring the magnetic properties via SQUID, the magnetization signal is nearly linear and the paramagnetic contribution is much smaller (Figure 3b). This further confirms that there is a large paramagnetic contribution from the GaSe base layer for our 1 ML MnSe$_2$ sample, shown in Figure 2.

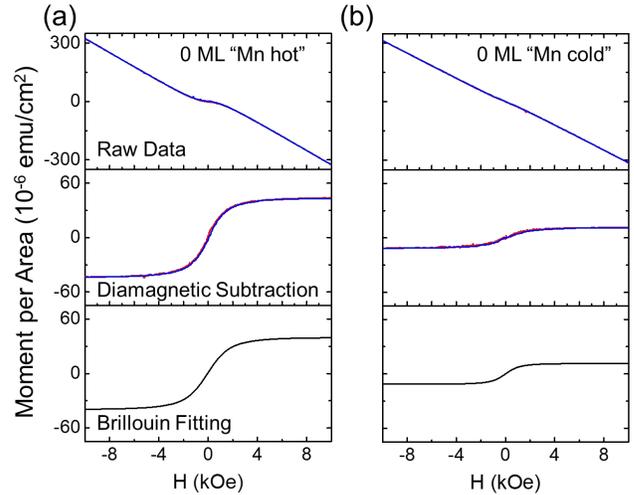

**Figure 3. 0 ML MnSe$_2$ magnetization control measurements with standard linear background subtraction and Brillouin function fitting. (a)** M(H) measurement of 0 ML "Mn hot" control sample using $T_{Mn}$ = 800°C. **(b)** M(H) measurement of 0 ML "Mn cold" control sample using $T_{Mn}$ = 620°C.

The control experiment of the "0 ML MnSe$_2$" sample shows that to accurately determine the ferromagnetic signal in our MnSe$_2$/GaSe(0001) samples, one has to also consider the paramagnetic contribution from the substrate. We apply a new method for background subtraction to the 1 ML MnSe$_2$ sample discussed in Figure 2. We proceed with the following equation to fit the non-hysteretic part of the SQUID data ($|H| \geq H_{cut}$),

$$m(H) = A_{dia} \cdot H + A_{para} \cdot B_{5/2}\left(\frac{\frac{5}{2}g^*\mu_B}{k_B T} \cdot H\right) + sign(H) \cdot m_{sat} \quad (1)$$

where $A_{dia}, A_{para}, g^*,$ and $m_{sat}$ are fitting parameters. The first term is a diamagnetic background that is linear in $H$. The second term is a paramagnetic background described by a Brillouin function with $J = 5/2$ for Mn, and $g^*$ is an effective g-factor observed in dilute magnetic semiconductors [Gaj 1979]. The last term is the saturated ferromagnetic magnetization, which adds as a positive/negative offset ($m_{sat}$) depending on the direction of the applied magnetic field.

To determine the magnetic moment per formula unit in the 1 ML MnSe$_2$ sample, we proceed with this new fitting method. Figure

4a shows the raw SQUID measurement for this sample. For the new method of background subtraction, we select a cutoff field, $H_{cut}$, that bounds the range of the ferromagnetic hysteresis loop. To demonstrate, we proceed with $H_{cut} = 2\ kOe$ and fit the data with Equation 1. After the fit is completed, the first two terms of Equation 1 (i.e. the diamagnetic and paramagnetic contributions) are subtracted from the raw data to yield the ferromagnetic hysteresis loop, shown in Figure 4b. Figure 4c shows the raw data, diamagnetic component, paramagnetic component, and ferromagnetic component over the field range of $\pm 5\ kOe$. This yields a value for $m_{sat}$ of 4.7 $\mu_B$/Mn which is in agreement with theoretical calculations [Kan 2014].

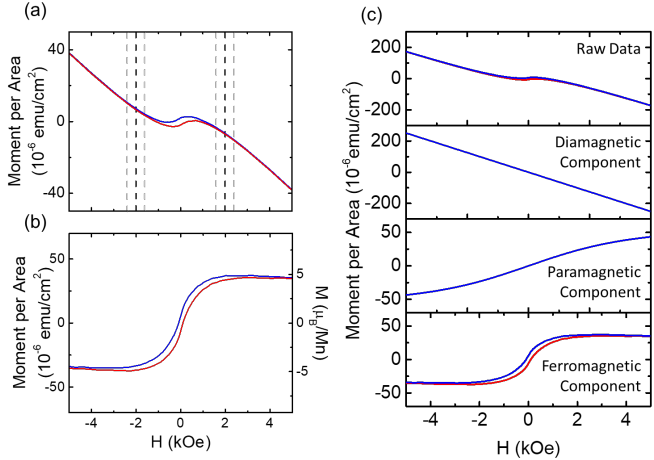

**Figure 4.** 1 ML MnSe$_2$ magnetic hysteresis loop using new background subtraction method considering diamagnetic and paramagnetic components using a Brillouin function background subtraction method. (a) Raw M(H) loop showing fitting parameters (dashed lines) and (b) final ferromagnetic hysteresis loop after background subtraction. (c) Background subtraction components (raw data, diamagnetic, paramagnetic, and ferromagnetic contributions) in 1 ML MnSe$_2$ grown on a 55 nm GaSe/GaAs(111) substrate.

The result of the fitting process depends on the choice of $H_{cut}$, which is determined by the field value that the hysteresis loop in the SQUID signal closes. An uncertainty in choosing $H_{cut}$ could in principle affect the extracted ferromagnetic signal. To show the robustness of the background subtraction method, we further performed the same fitting process with a few different $H_{cut}$ values, as shown in Table 1. The change of the $H_{cut}$ indeed changes the extracted magnetic moment in both the paramagnetic and ferromagnetic component. However, the total change of both components is within 20%, which indicate that the fitting method is accurate in determining the ferromagnetic response of the van der Waals magnet MnSe$_2$.

**Table 1.** Detailed description of Brillouin fitting parameters to determine magnetic moment in 1 ML MnSe$_2$ and 0 ML MnSe$_2$ control samples.

| $H_{cut}$ (kOe) | $M_{PM}$ (10$^{-6}$ emu/cm$^2$) | $M_{FM}$ (10$^{-6}$ emu/cm$^2$) | $M_{FM}$ ($\mu_B$/Mn) |
|---|---|---|---|
| 1.8 | 56 | 31 | 4.1 |
| 2.0 | 52 | 36 | 4.7 |
| 2.2 | 49 | 39 | 5.1 |

## IV. CONCLUSION

In summary, we have demonstrated that unintentional Mn doping of the GaSe(0001) base layer can introduce a strong paramagnetic signal in the MBE grown van der Waals magnet MnSe$_2$. This provides the physical basis for a new method of background subtraction that separates out diamagnetic and paramagnetic contributions from the ferromagnetic signal of monolayer MnSe$_2$. This is shown to be a reliable method for extracting the ferromagnetic hysteresis loop and value of the magnetic moment and should be considered for future studies of van der Waals magnets grown by MBE. Our study further shows the importance of performing control experiments on the magnetic background signal for epitaxially grown van der Waals magnets on a case-by-case basis. Further studies including comparisons with complementary techniques that are less sensitive to background signals, such as magneto-optic Kerr effect or anomalous Hall effect, will also be beneficial for accurately determining the magnetic moments in such systems.

## ACKNOWLEDGMENT

This work was supported by the U.S. Department of Energy under Grant DE-SC0018172. We also acknowledge the National Science Foundation MRI program under Grant DMR-1429143 for the MBE system. D.J.O. acknowledges the GEM National Consortium Ph.D. Fellowship.

## REFERENCES

Ataca C, Şahin H, Ciraci S (2012), "Stable, Single-Layer MX$_2$ Transition-Metal Oxides and Dichalcogenides in a Honeycomb-Like Structure," *The Journal of Physical Chemistry C,* vol. 116, pp. 8983-8999, doi: 10.1021/jp212558p.

Bonilla M, Kolekar S, Ma Y, Diaz H C, Kalappattil V, Das R, Eggers T, Gutierrez H R, Phan M-H, Batzill M (2018), "Strong room-temperature ferromagnetism in VSe$_2$ monolayers on van der Waals substrates," *Nature Nanotechnology*, doi: 10.1038/s41565-018-0063-9.

Deng Y, Yu Y, Song Y, Zhang J, Wang N Z, Wu Y Z, Zhu J, Wang J, Chen X H, Zhang Y (2018), "Gate-tunable Room-temperature Ferromagnetism in Two-dimensional Fe$_3$GeTe$_2$," *arXiv:1803.02038*.

Fei Z, Huang B, Malinowski P, Wang W, Song T, Sanchez J, Yao W, Xiao D, Zhu X, May A, Wu W, Cobden D, Chu J-H, Xu X (2018), "Two-Dimensional Itinerant Ising Ferromagnetism in Atomically thin Fe$_3$GeTe$_2$," *arXiv:1803.02559*.

Gaj J A, Planel R, Fishman G (1979), "Relation of magneto-optical properties of free excitons to spin alignment of Mn$^{2+}$ ions in Cd$_{1-x}$Mn$_x$Te," *Solid State Communications,* vol. 29, pp. 435-438, doi: https://doi.org/10.1016/0038-1098(79)91211-0.

Gong C, Li L, Li Z, Ji H, Stern A, Xia Y, Cao T, Bao W, Wang C, Wang Y, Qiu Z Q, Cava R J, Louie S G, Xia J, Zhang X (2017), "Discovery of intrinsic ferromagnetism in two-dimensional van der Waals crystals," *Nature,* vol. 546, p. 265, doi: 10.1038/nature22060.

Huang B, Clark G, Navarro-Moratalla E, Klein D R, Cheng R, Seyler K L, Zhong D, Schmidgall E, McGuire M A, Cobden D H, Yao W, Xiao D, Jarillo-Herrero P, Xu X (2017), "Layer-dependent ferromagnetism in a van der Waals crystal down to the monolayer limit," *Nature,* vol. 546, pp. 270-273, doi: 10.1038/nature22391.

Huang B, Clark G, Klein D R, MacNeill D, Navarro-Moratalla E, Seyler K L, Wilson N, McGuire M A, Cobden D H, Xiao D, Yao W, Jarillo-Herrero P, Xu X (2018), "Electrical control of 2D magnetism in bilayer CrI$_3$," *Nature Nanotechnology*, doi: 10.1038/s41565-018-0121-3.

Jiang S, Li L, Wang Z, Mak K F, Shan J (2018a), "Controlling magnetism in 2D $CrI_3$ by electrostatic doping," *Nature Nanotechnology*, doi: 10.1038/s41565-018-0135-x.

Jiang S, Shan J, Mak K F (2018b), "Electric-field switching of two-dimensional van der Waals magnets," *Nature Materials,* vol. 17, pp. 406-410, doi: 10.1038/s41563-018-0040-6.

Kan M, Adhikari S, Sun Q (2014), "Ferromagnetism in $MnX_2$ (X = S, Se) monolayers," *Physical Chemistry Chemical Physics,* vol. 16, pp. 4990-4994, doi: 10.1039/C3CP55146F.

Kim H H, Yang B, Patel T, Sfigakis F, Li C, Tian S, Lei H, Tsen A W (2018), "One million percent tunnel magnetoresistance in a magnetic van der Waals heterostructure," *arXiv:1804.00028*.

Klein D R, MacNeill D, Lado J L, Soriano D, Navarro-Moratalla E, Watanabe K, Taniguchi T, Manni S, Canfield P, Fernández-Rossier J, Jarillo-Herrero P (2018), "Probing magnetism in 2D van der Waals crystalline insulators via electron tunneling," *Science*, doi: 10.1126/science.aar3617.

Ma Y, Dai Y, Guo M, Niu C, Zhu Y, Huang B (2012), "Evidence of the Existence of Magnetism in Pristine $VX_2$ Monolayers (X = S, Se) and Their Strain-Induced Tunable Magnetic Properties," *ACS Nano,* vol. 6, pp. 1695-1701, doi: 10.1021/nn204667z.

Mermin N D, Wagner H (1966), "Absence of Ferromagnetism or Antiferromagnetism in One- or Two-Dimensional Isotropic Heisenberg Models," *Physical Review Letters,* vol. 17, pp. 1133-1136.

O'Hara D J, Zhu T, Trout A H, Ahmed A S, Luo Y K, Lee C H, Brenner M R, Rajan S, Gupta J A, McComb D W, Kawakami R K (2018), "Room Temperature Intrinsic Ferromagnetism in Epitaxial Manganese Selenide Films in the Monolayer Limit," *Nano Letters,* vol. 18, pp. 3125-3131, doi: 10.1021/acs.nanolett.8b00683.

Onsager L (1944), "Crystal Statistics. I. A Two-Dimensional Model with an Order-Disorder Transition," *Phys. Rev.,* vol. 65, pp. 117-149.

Pekarek T M, Crooker B C, Miotkowski I, Ramdas A K (1998), "Magnetic measurements on the III-VI diluted magnetic semiconductor $Ga_{1-x}Mn_xSe$," *Journal of Applied Physics,* vol. 83, pp. 6557-6559, doi: 10.1063/1.367781.

Sivadas N, Daniels M W, Swendsen R H, Okamoto S, Xiao D (2015), "Magnetic ground state of semiconducting transition-metal trichalcogenide monolayers," *Phys. Rev. B,* vol. 91, p. 235425.

Song T, Cai X, Tu M W-Y, Zhang X, Huang B, Wilson N P, Seyler K L, Zhu L, Taniguchi T, Watanabe K, McGuire M A, Cobden D H, Xiao D, Yao W, Xu X (2018), "Giant tunneling magnetoresistance in spin-filter van der Waals heterostructures," *Science*, doi: 10.1126/science.aar4851.

Walsh L A, Hinkle C L (2017), "van der Waals epitaxy: 2D materials and topological insulators," *Applied Materials Today,* vol. 9, pp. 504-515, doi: https://doi.org/10.1016/j.apmt.2017.09.010.

Wang Z, Gutiérrez-Lezama I, Ubrig N, Kroner M, Gibertini M, Taniguchi T, Watanabe K, Imamoğlu A, Giannini E, Morpurgo A F (2018a), "Very large tunneling magnetoresistance in layered magnetic semiconductor $CrI_3$," *Nature Communications,* vol. 9, p. 2516, doi: 10.1038/s41467-018-04953-8.

Wang Z, Sapkota D, Taniguchi T, Watanabe K, Mandrus D, Morpurgo A F (2018b), "Tunneling Spin Valves Based on $Fe_3GeTe_2$/hBN/$Fe_3GeTe_2$ van der Waals Heterostructures," *Nano Letters*, doi: 10.1021/acs.nanolett.8b01278.

Wang Z, Zhang T-Y, Ding M, Dong B, Li Y-X, Chen M-L, Li X-X, Li Y, Li D, Jia C-K, Sun L-D, Guo H, Sun D-M, Chen Y-S, Yang T, Zhang J, Ono S, Han Z V, Zhang Z-D (2018c), "Electric-field control of magnetism in a few-layered van der Waals magnet," *arXiv:1802.06255*.

Wenyu X, Yangyang C, Patrick M O, Xiao Z, Wei Y, Tang S, Qi S, Tianyu W, Jiangnan Z, Shuang J, Xie X C, Yan L, Wei H (2017), "Electric field effect in multilayer $Cr_2Ge_2Te_6$ : a ferromagnetic 2D material," *2D Materials,* vol. 4, p. 024009.

Zhong D, Seyler K L, Linpeng X, Cheng R, Sivadas N, Huang B, Schmidgall E, Taniguchi T, Watanabe K, McGuire M A, Yao W, Xiao D, Fu K-M C, Xu X (2017), "Van der Waals engineering of ferromagnetic semiconductor heterostructures for spin and valleytronics," *Science Advances,* vol. 3.